# Melting of charge density waves in low dimensions

## Graphical abstract

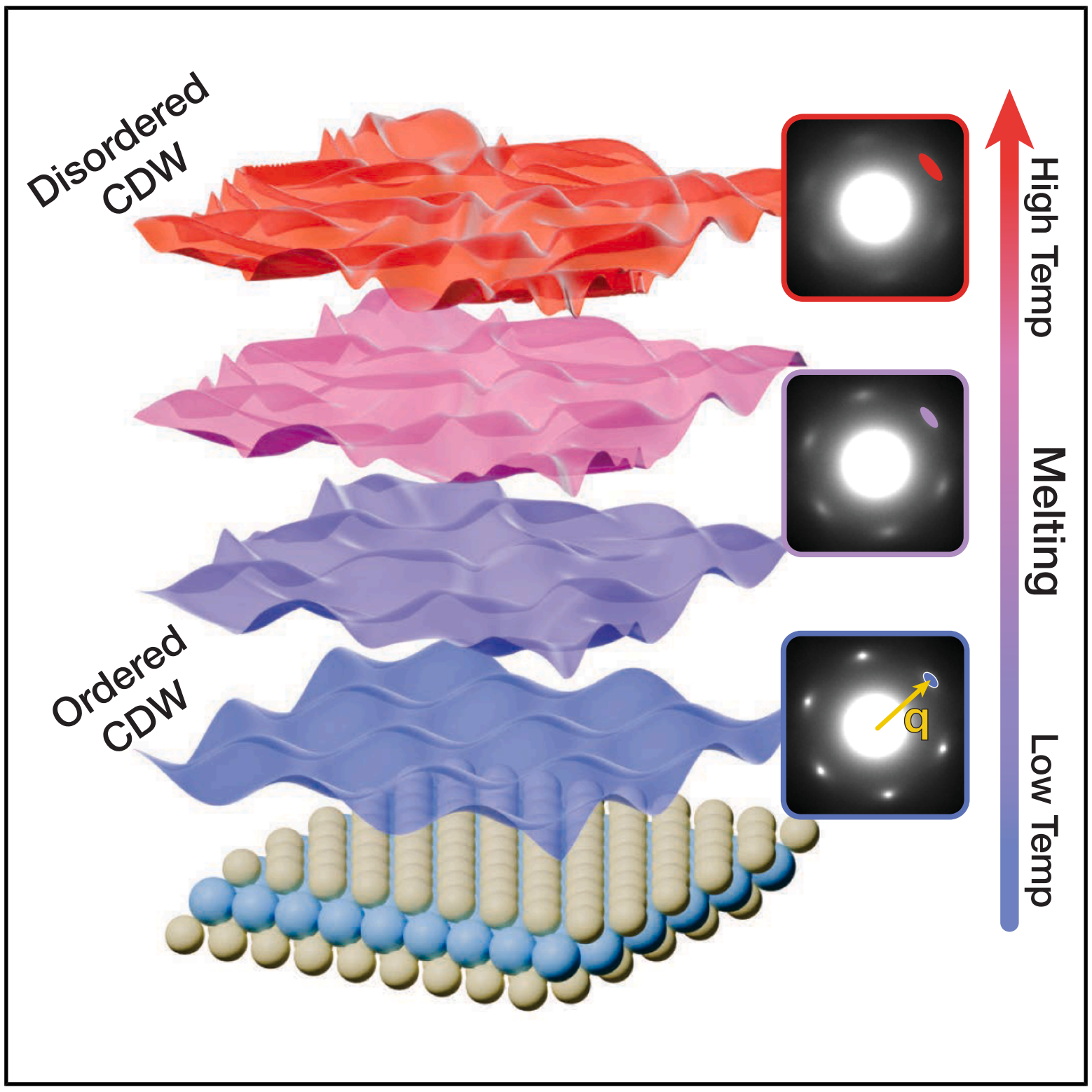

## Highlights

- Two-dimensional CDWs melt hexatically
- CDW wave vector contracts and azimuthally broadens during melting
- CDW amplitude diminishes during melting, a process unique from classical melting
- CDW melting is a dislocation-driven process

## Authors

Jeremy M. Shen, Alex Stangel, Suk Hyun Sung, ..., Ismail El Baggari, Kai Sun, Robert Hovden

## Correspondence

hovden@umich.edu

## In brief

Charge density waves melt continuously through partially ordered nematic and hexatic states. As a charge density wave is thermally excited, disorder emerges progressively—initially through smooth elastic deformations that modulate the local wavelength and subsequently via the nucleation of dislocations. Across material systems, there are three salient signatures of electronic melting in low dimensions: azimuthal superlattice peak broadening, wavevector contraction, and integrated intensity decay.

Article

# Melting of charge density waves in low dimensions

Jeremy M. Shen,[1,6] Alex Stangel,[2,6] Suk Hyun Sung,[3] Nishkarsh Agarwal,[2] Gaihua Ye,[4] Cynthia Nnokwe,[4] Liuyan Zhao,[5] Yang Zhang,[3] Rui He,[4] Ismail El Baggari,[3] Kai Sun,[5] and Robert Hovden[2,5,7,*]
[1]Department of Electrical and Computer Engineering, University of Michigan, Ann Arbor, MI 48109, USA
[2]Department of Materials Science and Engineering, University of Michigan, Ann Arbor, MI 48109, USA
[3]The Rowland Institute at Harvard, Cambridge, MA 02138, USA
[4]Department of Electrical and Computer Engineering, Texas Tech University, Lubbock, TX 79409, USA
[5]Department of Physics, University of Michigan, Ann Arbor, MI 48109, USA
[6]These authors contributed equally
[7]Lead contact
*Correspondence: hovden@umich.edu

**PROGRESS AND POTENTIAL** Here, we show that incommensurate charge density waves (CDWs) undergo a melting process, similar to classical melting of solids, but with distinct signatures. Under applied heat, CDWs fluctuate through elastic deformation and dislocations that degrade long range order, cause characteristic expansion of the CDW wavelength, and diminish the CDW amplitude. In lower dimensions, CDW melting is even more exotic—progressing through an intermediate phase where orientational order is retained despite a loss of translational symmetry. This all takes place while the underlying atomic lattice remains unchanged. Here, a two-dimensional CDW in endotaxial $TaS_2$ at elevated temperatures provides a model system to elucidate topologically driven CDW melting. However, a meta-analysis of previously reported CDW systems—including $TaSe_2$, BSCMO, LSCO, and $TbTe_3$—suggests that continuous CDW melting processes are present across materials. Computational, theoretical, and experimental methods show agreement. More broadly, this work highlights the macroscopic influence local disorder and dislocations have in quantum materials.

## SUMMARY

Charge density waves (CDWs) are collective electronic states that can reshape and melt, even while confined within a rigid atomic crystal. In two dimensions, melting is predicted to be distinct, proceeding through partially ordered nematic and hexatic states that are neither liquid nor crystal. Here, we measure and explain how continuous, hexatic melting of incommensurate CDWs occurs in low-dimensional materials. As a CDW is thermally excited, disorder emerges progressively—initially through smooth elastic deformations that modulate the local wavelength and subsequently via the nucleation of topological defects. Experimentally, we track three hallmark signatures of CDW melting—azimuthal superlattice peak broadening, wavevector contraction, and integrated intensity decay.

## INTRODUCTION

Thermal melting of crystals is a dramatic transformation. It is governed by the emergence of topological defects—dislocations that tear the underlying crystalline order. As dislocations proliferate, they disrupt the periodic lattice, driving volumetric expansion and loss of long-range order.[1,2] Over the past century, it has become clear that melting is dictated by microstructural disorder and dimensionality.[3–5] In two dimensions, melting can proceed via exotic intermediate phases, such as hexatic and nematic order, that retain orientational coherence despite the loss of translational symmetry—offering partially ordered states that interpolate between a crystalline and liquid phase.[6] Similar hexatic and nematic order of electronic, quantum, and magnetic states are predicted in low-dimensional materials.[7]

Charge density waves (CDWs) represent an electronic analog to classical crystallinity, spontaneously breaking translational symmetry to form periodic modulations in the electronic charge density. These emergent electronic states are often intertwined with other collective phenomena such as superconductivity and Mott insulating behavior and have been observed in materials ranging from transition metal dichalcogenides (TMDs) to cuprates and manganites.[8–10] Like atomic lattices, CDWs elastically deform and contain topological defects—including dislocations and domain boundaries—that destabilize order.[11] Electron microscopy has directly imaged such CDW dislocations in

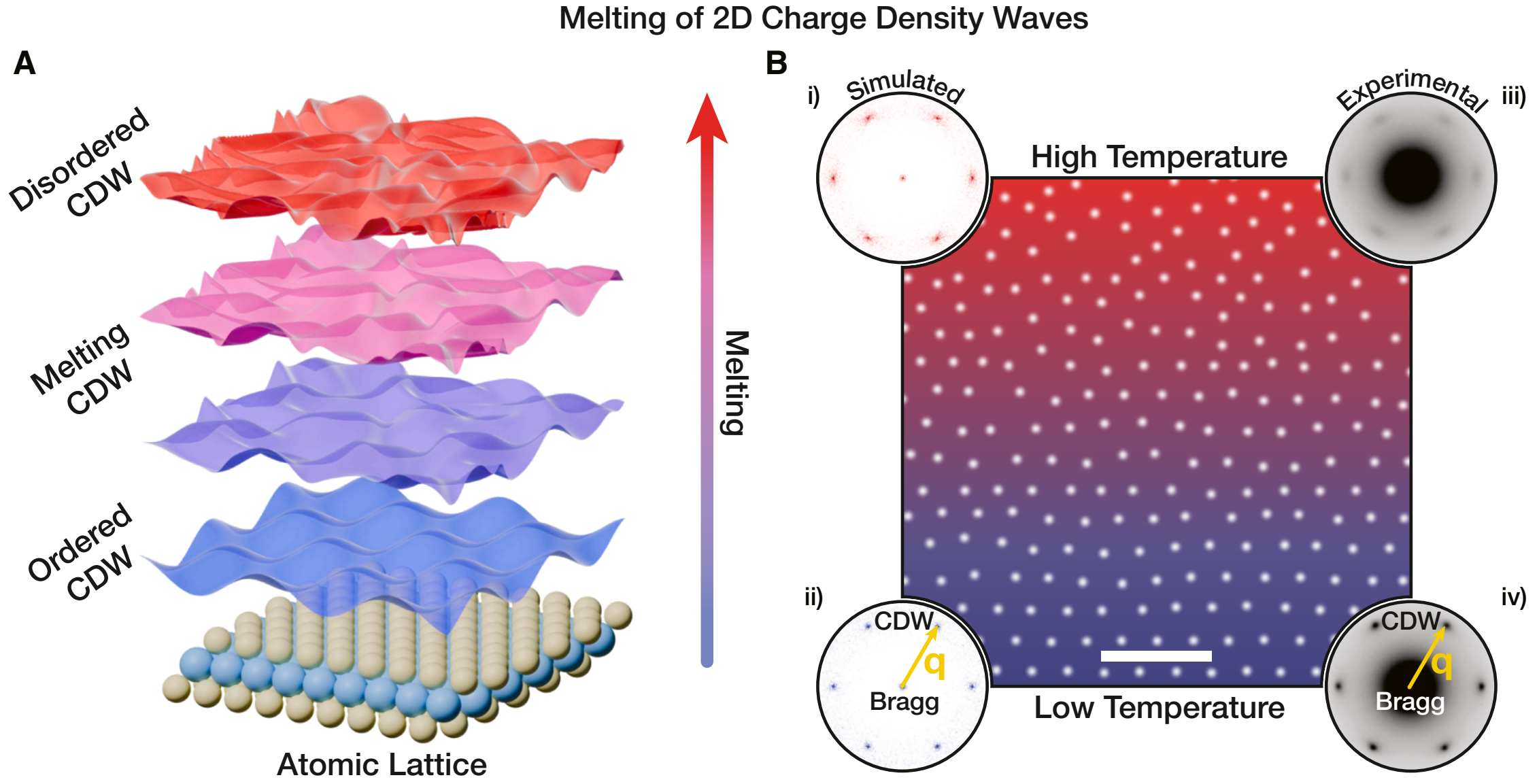

**Figure 1. Melting of two-dimensional charge density waves**

(A) Illustration of a progressively melted 2D CDW. The CDW becomes less spatially coherent, melting independently of the underlying atomic lattice.

(B) Melting of CDWs modeled with a lattice of CDW peaks (white dots). Scale bar, 5 nm. At higher temperatures, CDW fluctuations become large and disordered. Comparison of experimental 2D 1T-$TaS_2$ CDW diffraction (iii and iv) with simulated CDW diffraction (i and ii) at low and high temperatures. Molecular dynamic simulation of CDW melting qualitatively matches experimental electron diffraction.

pristine crystalline regions of complex oxides. Kourkoutis and colleagues have shown that CDWs incommensurate with the crystal lattice exhibit significant local disorder.[12] Similar dislocations of the CDW have been reported in the surface states of $TaS_2$[13,14] and transient defects of phonon dynamics in 1T-$TiSe_2$.[15] Recently, the creation of CDW dislocations has been observed directly under applied heat.[16]

The thermal creation of CDW dislocations is expected to drive the melting of CDWs, resulting in a loss of long-range CDW order. In two-dimensional 1T-$TaS_2$, hexatic CDW melting was recently reported[17]; however, the accompanying microstructural evolution—and particularly the observed contraction of the CDW wavevector **q**—remains unexplained and seems inconsistent with classical theories of melting. CDW melting occurs at temperatures far below the atomic melting point of the host lattice.

Here, we describe melting of incommensurate CDWs. We show that fluctuations of CDW order—through elastic deformation and dislocations—cause characteristic expansion of the CDW with hexatic order. To accommodate CDW expansion within a fixed volume, there is a concomitant reduction in CDW amplitude. This melting of CDWs has three characteristics visible in diffraction: azimuthal broadening of the wavevector due to hexatic/nematic disorder, contraction of the wavevector due to a reduced packing density of the charge crystal, and a reduction of total diffracted intensity due to attenuation of the CDW amplitude. High-temperature CDWs in two-dimensional 1T-$TaS_2$ serve as a model system for CDW melting with clear connection between experiment, computational, and Landau descriptions. A meta-analysis of 28 previously reported incommensurate CDW systems—including $TaSe_2$, BSCMO, LSCO, and $TbTe_3$—suggests that continuous CDW melting processes are broadly present across materials.

## RESULTS

### Hexatic melting of 2D charge crystals

The CDW is a periodic modulation of valence electron density within the crystal (Figure 1A) and intimately couples to the atomic lattice through electron-phonon interactions.[18,19] Electron diffraction is sensitive to atomic movement that directly corresponds to CDW structure. Figure 1B illustrates the CDW structure across temperature. At low temperatures, the CDW is ordered (Figure 1B, bottom) with a well-defined wavevector (**q**) (Figure 1B, ii and iv). As the CDW is thermally excited, disorder emerges progressively—initially through smooth elastic deformations that modulate the local wavelength and subsequently via the nucleation of topological defects that more severely disrupt long-range coherence (Figure 1B, top).

There are three signatures for hexatic/nematic melting of CDWs that are salient in diffraction: (1) superlattice peaks associated with the CDW blur azimuthally, (2) a shortening of the principal CDW wavevector, and (3) a decreasing integrated CDW intensity (Figure 2). The first two features are known to classical melting of crystal solids.[20] The third feature is unique to melting of CDWs.

**Blurring of CDW peaks:** The primary feature of hexatic melting in diffraction is azimuthal blurring of peaks.[21–25] Sharp peaks and diffuse rings describe the temperature extremes of ordered and amorphous CDW structures, respectively. *In situ* electron diffraction of 2D 1T-$TaS_2$ (Figure 2B) shows the evolution of an incommensurate CDW superlattice peak as it melts (A and B) (A) Experimental line profile through CDW superlattice peaks in (B). The CDW peaks are initially sharp (iii) but blur azimuthally and decrease in intensity continuously (i and ii). Black line tracking **q**-vector is a guide to the eye.

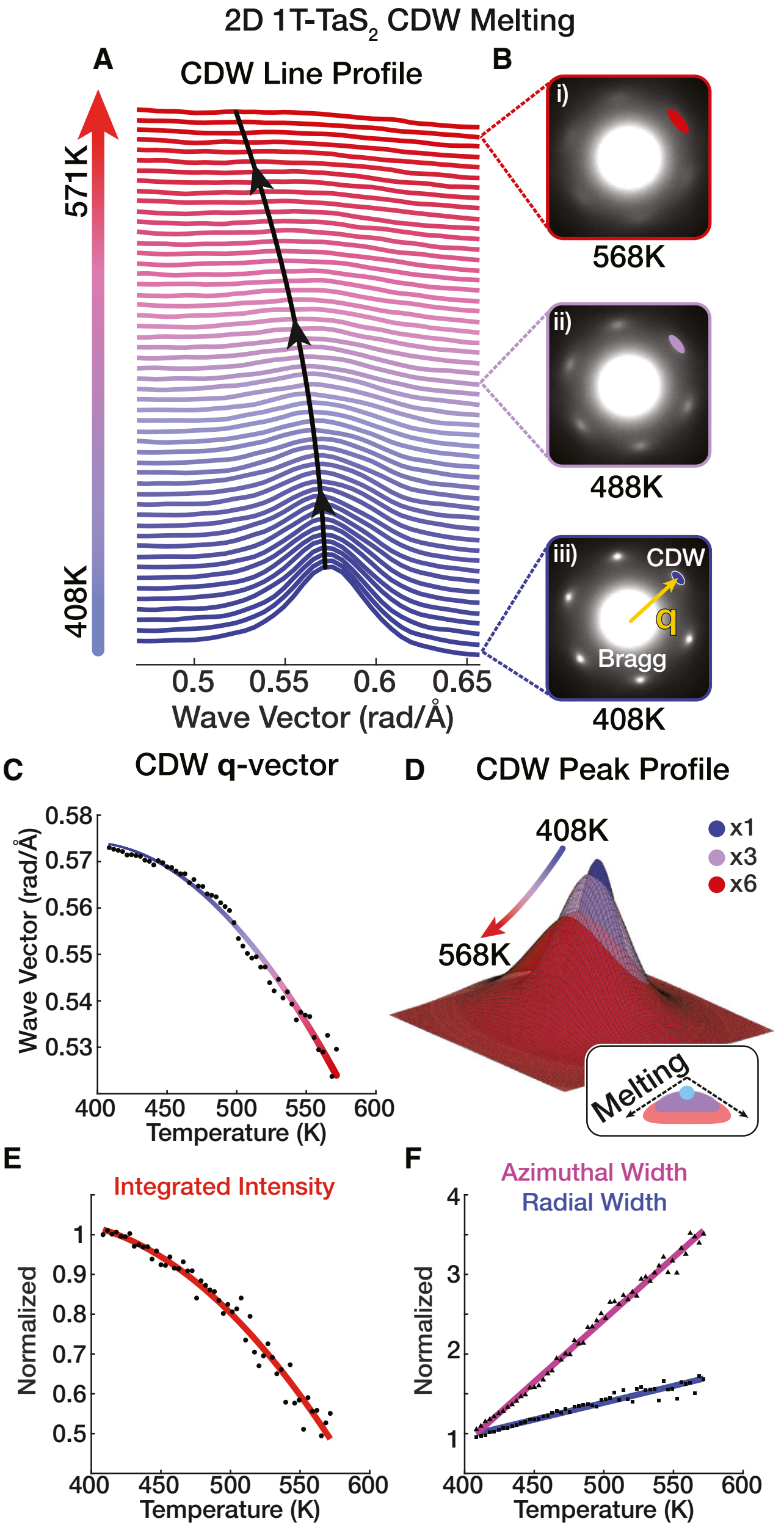

**Figure 2. Evolution of CDW melting in diffraction (2D 1T-$TaS_2$)**

(C) The CDW **q**-vector—determined by the peak maximum location—contracts by over 8.6% from 408 to 571 K.

(D–F) (D) Superlattice peak profiles at 408 K (blue), 488 K (purple), and 568 K (red) scaled by 1×, 3×, and 6×, respectively. (Inset) Cartoon depiction of superlattice peak evolution: peak position shifting to lower wavenumbers and peak blurring. Through the melting regime, (E) the integrated intensity halves and (F) the azimuthal and radial width increases 3.5× and 1.7×, respectively. These features of CDW melting evolve concurrently.

continuously when heated from 408 to 571 K. Initially, there are six sharp peaks, but upon melting, the peaks begin to blur into the background, becoming more diffuse with increasing temperature. At 571 K, the CDW peak in $TaS_2$ is 3.5 times broader along the azimuth and 1.7 times broader radially than at 408 K (Figure 2F).

In lower dimensions, translational symmetry is lost before orientational order, a sequence described by the KTHNY theory.[3,6,26,27] In this continuous progression of the hexatic phase, diffraction peaks broaden azimuthally as translational order decays, while local 6-fold coordination—and thus orientational order—persists (Figure 1B, i and iii). Six first-order CDW peaks that are radially sharp but azimuthally broad are uniquely attributed to the intermediary hexatic state.

Alongside simulated diffraction patterns of the CDW (Figure 1B, i and ii) are experimental diffraction patterns (Figure 1B, iii and iv) taken from 2D 1T-$TaS_2$ at 408 and 568 K. In real space, we see how the proliferation of disorder in the charge crystal destroys spatial coherence in the charge density, resulting in blurred superlattice peaks (Figure S1). Eventually, the state becomes a fully melted liquid with short-range order. In $TaS_2$, a complete liquid state was observed using ultrafast electron diffraction by Kogar and colleagues.[28]

CDW melting is illustrated in Figure 1B using a temperature-dependent molecular dynamics simulation of 2D CDWs (see Methods). Melting is computationally modeled by allowing charge density centers to thermally move according to Boltzmann statistics. Charge centers interact through a divergent short-range repulsion—mimicking Pauli exclusion—and a weaker long-range attraction that binds the modulation. The system evolves through an iterative Monte Carlo process until convergence at each temperature.

**CDW wavevector contraction:** During hexatic or nematic melting, the CDW **q**-vector contracts due to CDW expansion caused by disorder. Figure 2 experimentally shows the shift toward lower wavevectors in the CDW of 2D 1T-$TaS_2$. The CDW **q**-vector contracts continuously over the temperature regime (408–571 K) (Figure 2C). The **q**-vector contraction happens simultaneously with the peak blurring and decreasing integrated intensity. Figure 2 shows the peak evolution of the measured diffraction pattern, including the background—as often reported. At high temperatures nearing the liquid phase, the rate of **q**-vector contraction appears to accelerate. However, above roughly 450 K in $TaS_2$, the non-uniform background exaggerates the **q**-vector contraction of weak and broad CDW peaks. In $TaS_2$, we report a q-vector contraction of 2% when heated from 408 to 571 K after background subtraction (Figure S5).

The atomic lattice remains relatively unchanged throughout the CDW melting process. Over the CDW melting regime, previously reported CDW expansion is roughly 10–500 times larger than thermal expansion in the host atomic crystal.[29–33] For 1T-$TaS_2$, the CDW expansion is around 7.5 times larger than that of the crystal lattice.[34] Furthermore, CDW transitions occur at temperatures far below the atomic melting point of the host lattice and thus within a nearly constant atomic volume.

The proliferation of CDW dislocations pushes apart the CDW lattice forcing expansion in real space (**q**-vector contraction). When atomic crystals melt, the average lattice spacing increases

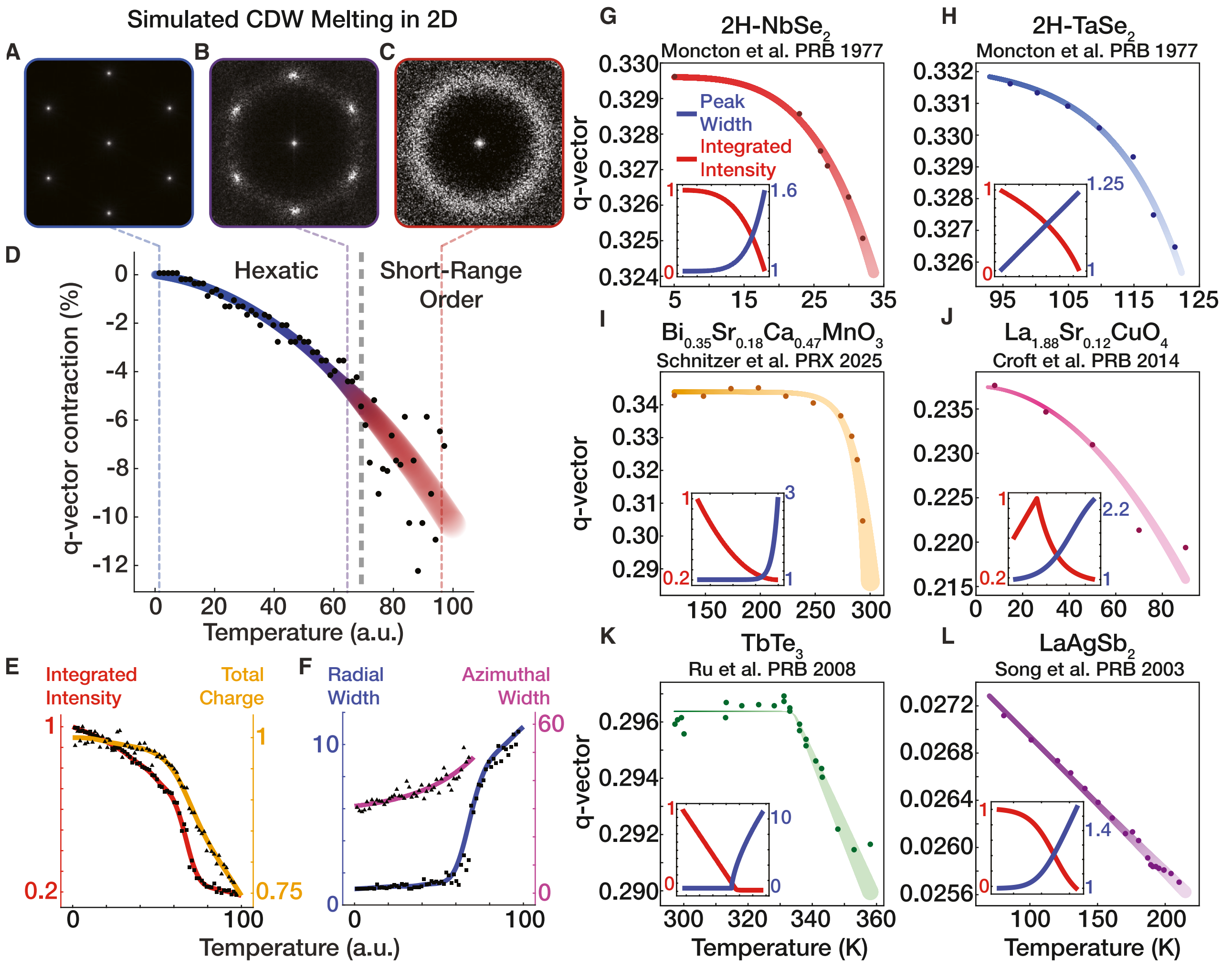

**Figure 3. Melting of incommensurate CDWs in previously reported literature and a simulated CDW**

(A–C) Simulated diffraction patterns of melted charge lattice with (D) wavevector contraction, (E) integrated intensity decay, and (F) peak width broadening. (E) The number of CDW peaks decreases with melting. (F) The superlattice peaks blur azimuthally in the hexatic regime before becoming amorphous rings (purple). There is slight radial broadening in the hexatic regime that rapidly accelerates as it becomes amorphous (blue). Wavevector contraction, peak broadening, and integrated intensity decrease quantified for (G and H) 2D transition metal dichalcogenides,[35] (I) manganites,[16] (J) cuprates,[36] (K) rare-earth tellurides,[31] and (L) 2D metals.[37] Fitted lines are guides to the eye, where line width and color intensity are proportional to the CDW peak width and integrated intensity, respectively. Many more material systems (see Figures S6–S8) undergo a continuous melting process for incommensurate CDWs.

and the total volume expands. In contrast, CDWs melt within a rigid crystal volume well below the host-lattice melting point. A fully classical treatment of CDW melting in a constant volume unphysically drives up electronic pressure that compresses the CDW (**q**-vector expansion) (Figure S2). Thus, classical melting in a fixed volume contradicts the observed CDW melting behavior where the CDW lattice expands (**q**-vector contracts). For CDWs, the total amplitude of charge density centers need not be conserved. A description of CDW melting requires a spatially fluctuating charge density amplitude—akin to a grand canonical ensemble.

**CDW intensity weakens:** CDW melting departs from classical solid melting, proceeding with the simultaneous emergence of disorder and localized CDW amplitude collapse. A local CDW collapse around defect centers relieves electronic pressure without causing volumetric expansion of the CDW. This effectively lowers the number of charge density peaks as the temperature rises and the average real space wavelength increases (**q** contracts), while the system minimizes its free energy within a constant volume (Figure S2).

The decreasing integrated intensity of CDW superlattice peaks is a direct measurement of CDW attenuation during melting. In $TaS_2$, the integrated peak intensity halves over the temperature regime (408–571 K) (Figure 2E). This rate of decrease far outpaces the thermal disorder effect on Bragg peaks (Figure S3). Simulations closely match experimental diffraction of the continuous contraction of the incommensurate CDW wavevector and the decay of the peak's integrated intensity (Figures 3A–3F). Note, the integrated intensity is distinct from the CDW peak height, which will decrease as it broadens.

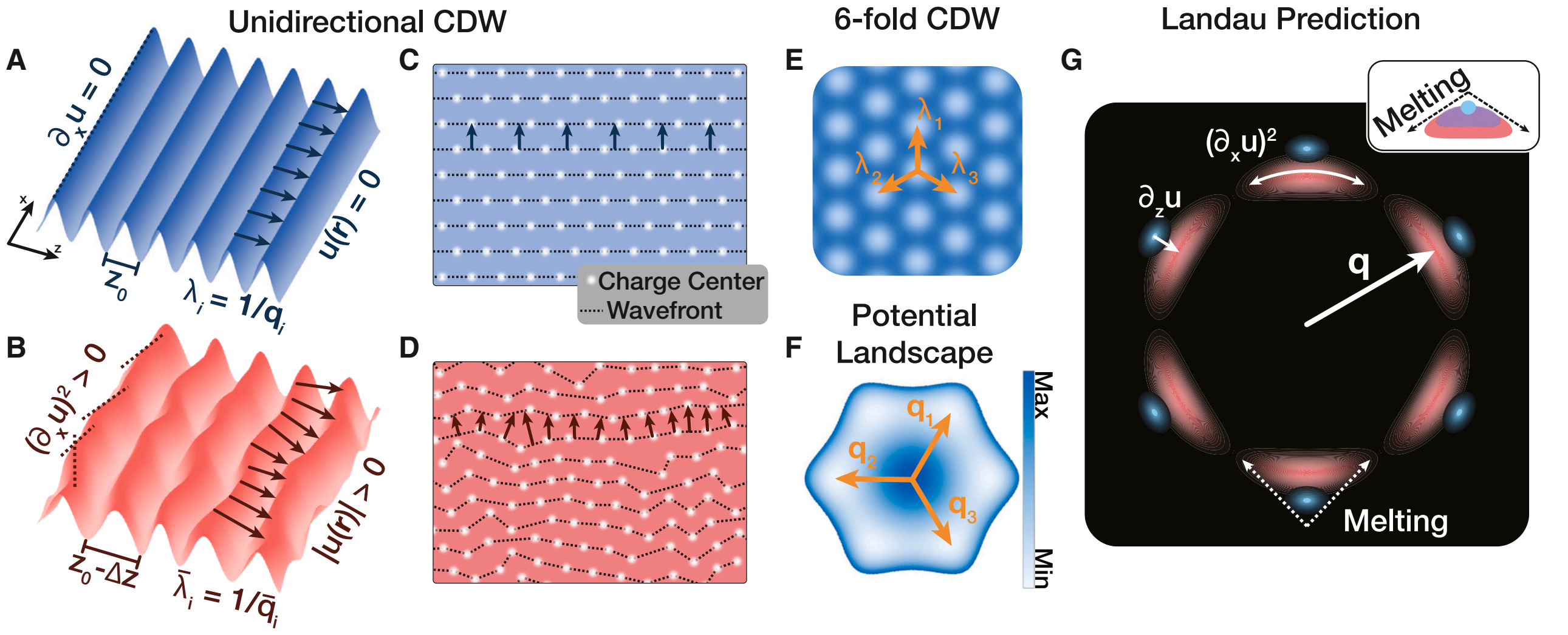

**Figure 4. Fluctuation-driven melting of charge density waves**

(A–E) Illustration of (A) ordered and (B) disordered unidirectional CDWs. Fluctuations in the wavefront $u(\mathrm{r})$ drive a wavelength increase by $-\Delta Z$. (C and D) In the charge lattice, planes of charge centers are discrete representations of the CDW wavefront. (E) Incommensurate CDW in 2D 1T-$TaS_2$ is a triple CDW system with 6-fold symmetry. Here, we use $\lambda_i$ to denote real space CDW wavelength.

(F) Potential landscape in reciprocal space constructed with Landau theory.

(G) Predicted peak evolution of wavefronts consistent with the Landau free energy landscape. Melting drives CDW wavevector contraction and peak broadening.

Temperature-dependent (77–600 K) Raman spectroscopy of 2D 1T-$TaS_2$ shows a continuous decrease in integrated intensity along with a large broadening of the 70 $cm^{-1}$ Raman peak (Figure S4). This suggests that no amplitude mode characteristic to the incommensurate CDW of 1T-$TaS_2$ has been detected by high-temperature Raman spectroscopy. Similar behavior has been observed in other CDW systems such as rare-earth tellurides,[38] 1T-$TaSe_2$,[39] 2H-$NbSe_2$,[40] and even bulk 1T-$TaS_2$.[41]

## Nearly universal CDW melting behavior

Characteristic continuous melting of incommensurate CDWs is observed in nearly all (quasi-) 2D CDW systems at elevated temperatures. We compile a meta-study of melting behavior of CDWs probed via X-ray, neutron, and electron diffraction (Figures 3G–3L and S6–S8). Incommensurate CDWs in bulk TMDs such as 2H-$TaSe_2$ and 2H-$NbSe_2$ melt similarly to 2D 1T-$TaS_2$.[35,42–44] We find this behavior occurs in 2-, 3-, and 4-fold in-plane crystal symmetries. Wavevector contraction, peak width increase, and integrated intensity decrease are observed in melting of incommensurate CDWs in manganites (BSCMO),[12,16,45–47] 2D metals such as $(K_{1-x}Rb_x)_3Cu_8S_6$ and $LaAgSb_2$,[37,48,49] rare earth tellurides $TbTe_3$ and $Gd_2Te_5$,[31,50] and other CDW systems such as $UPt_2Si_2$, $LaPt_2Si_2$, $CuV_2S_4$, and $U_2Ti$.[33,51–53] Croft et al. report these melting features in one cuprate system $La_{1.88}Sr_{0.12}CuO_4$.[36] A scanning tunneling microscopy study of the 2D surface CDW in $NbSe_3$ shows consistent melting behavior.[54]

However, it should be noted that the wavevector evolution of incommensurate CDWs in cuprates varies greatly with doping concentration and both contraction and expansion of the CDW have been reported.[36,55–57] No observable trend in wavevector evolution is present in $BaNi_2As_2$.[58] Melting of (quasi-)1D CDWs show no clear preference for wavevector contraction or expansion (Figures S8 and S9). $SmNiC_2$, $Ta_2NiSe_7$, and $Er_5Ir_4Si_{10}$ exhibit wavevector contraction,[59–61] while $Ho_5Ir_4Si_{10}$, blue bronzes, and $NbSe_3$ exhibit wavevector expansion.[62–65] Wavevector expansion is observed in orthorhombic $TaS_3$, while in the monoclinic polymorph, the wavevector remains unchanged.[66]

In this meta-study, the three signatures of CDW melting appear in nearly all incommensurate 2D CDWs across various crystal symmetries. Emphasis should be placed on qualitative trend analysis in regimes where CDW amplitudes become weak—in many cases, background fitting was not reported and detector efficiency was limited.

## Fluctuation-driven CDW expansion

The physical origin of CDW expansion can be understood by combining insights from two seminal works.[67,68] We show that disorder and fluctuations arising from CDW melting naturally lead to an expansion of the CDW wavelength.

To illustrate CDW wavelength expansion, we consider a unidirectional CDW with wavevector $\mathbf{q}$ and order parameter $\psi(\mathbf{r}) = Ae^{i\mathbf{q}\cdot\mathbf{r}}$ (Figures 4A and 4C). The theory readily generalizes to more complex cases, such as hexagonal CDWs formed by a superposition of three unidirectional components (Figure 4E). We define $\mathbf{r}=(x,z)$, with $x$ and $z$ denoting directions perpendicular and parallel to $\mathbf{q}$, respectively. Melting of the CDW introduces strong fluctuations and disorder, resulting in dislocations and distortions of the CDW wavefront (Figures 4B and 4D). These effects are captured by fluctuating the order parameter,

$$\psi(\mathbf{r}) = A(\mathbf{r})e^{iq(z+u(\mathbf{r}))}, \quad \text{(Equation 1)}$$

where $A(\mathbf{r})$ and $qu(\mathbf{r})$ describe the amplitude and phase fluctuations, respectively. Here we focus on fluctuating displacements of the wavefront, $u(\mathbf{r})$, that are prevalent in 2D CDW melting.

As pointed out by McMillan,[67] for an incommensurate CDW, the frequency of phase fluctuations vanishes in the long-wavelength limit ($k\rightarrow 0$), implying that, to leading order, the free energy includes only terms such as $(\partial_x u)^2$ and $(\partial_z u)^2$. While sufficient for many purposes, these terms alone cannot explain melt-induced CDW expansion where higher-order terms become essential. Fortunately, Grinstein and Pelcovits[68] derived such nonlinear contributions in their study of smectics and 1D solids. Combining their results with McMillan's theory, we obtain the effective free energy of a disordered CDW as follows:

$$H = \frac{1}{2}KA^2q_0^4 \int d^2r \left\{ C_0(\partial_x u)^2 + \frac{1}{2q_0^2}\left(\partial_x^2 u\right)^2 + 2(\partial_z u)^2 + 2\partial_z u(\partial_x u)^2 + \frac{1}{2}\left[(\partial_x u)^2\right]^2 + \text{higher order terms} \right\}. \quad \text{(Equation 2)}$$

The first three terms are quadratic in $u$ and describe linear responses; the remaining terms govern nonlinear behavior. In smectics (Grinstein and Pelcovits), continuous rotational symmetry prohibits the $(\partial_x u)^2$ term, enforcing $C_0 = 0$. However, for CDWs in crystals with discrete rotational symmetry (Figure 4F), a positive $C_0$ must be included to match McMillan's theory (see supplemental information); this term is symmetry allowed, and its positive sign is required due to stability. In addition, the presence of the lattice can also renormalize other coefficients, but we neglect such corrections at leading order.

The free energy contains two key deformation modes: $\partial_x u$, describing CDW wavefront distortions, and $\partial_z u$ controlling the wavevector magnitude—i.e., contraction ($\partial_z u > 0$) or expansion ($\partial_z u < 0$) of the CDW (corresponding to an expansion and contraction of **q**, respectively). At quadratic order, these modes are decoupled. However, the cubic term $\partial_z u(\partial_x u)^2$, the leading-order nonlinear term allowed by symmetry, couples them and plays a pivotal role. In smectics, this cubic coupling underlies fluctuation-driven renormalizations[68] and stretch-induced buckling instabilities.[69]

In the context of CDW melting, the nonlinear term plays a similarly crucial role. Melting introduces disorder and dislocations, leading to an increased $\langle(\partial_x u)^2\rangle$, which quantifies wavefront distortions. Due to the positive coefficient of this cubic coupling, a negative $\partial_z u$ lowers the free energy, indicating that wavevector contraction—and thus wavelength expansion—is energetically favored. As a result, crumpled CDW wavefronts due to the melting naturally induce an expansion of the CDW wavelength, as illustrated in Figure 4D. The peak evolution in Figure 4G is empirically constructed for consistency with the Landau model and experimental observations in $TaS_2$.

Amplitude fluctuations of the CDW become important around CDW dislocations. Outside of these regions, the CDW is more uniform and the amplitude changes are small.[12] This behavior can be captured through a typical Landau expansion of $\psi(\mathbf{r})$ where amplitude fluctuations are allowed. In this picture, the amplitude of the CDW couples directly to disorder in its phase as $A^2(r)(\nabla qu(\mathbf{r}))^2$. As the CDW structure becomes disordered, the amplitude of the CDW will decrease. In regions around dislocation cores, the phase term $(\nabla qu(\mathbf{r}))^2$ diverges causing total amplitude collapse.

## DISCUSSION

In summary, we show that 2D incommensurate CDWs melt continuously through intermediate hexatic and nematic phases. **q**-vector contraction and CDW amplitude reduction are an intrinsic feature of CDW dislocations introduced during the melting process. CDW melting is distinct from classical melting: the emergence of CDW dislocations drives the disappearance of local charge density that relieves electronic pressure, minimizes its free energy, and accommodates contraction of **q** within a fixed volume. Melting gives the appearance of an increase in incommensurability. However, long-range CDW order is not present at elevated temperatures and the incommensurate **q**-vector is tied to CDW fluctuations. This melting behavior is found in nearly all incommensurate 2D CDW systems with various crystal symmetries. The melting of incommensurate CDWs does not preclude the existence of first-order CDW phase transitions—rather the melting is a continuous process that progresses above these phase transitions. Often, CDWs exist in materials that host rich phases such as superconductivity, Mott insulating phases, and magnetic order. These competing correlation-driven orders become increasingly relevant at lower temperatures, and a complete physical model of disorder would be extended to capture such interactions. The work here focuses on CDW melting in higher-temperature regimes but does not exclude additional interactions. Melting is a process that highlights how subtle changes in diffraction manifest from complex microstructure in the real space order parameter of quantum states.

## METHODS

### Electron microscopy

*In situ* selected area electron diffraction (SAED) was performed on Thermo Fisher Scientific (TFS) Talos (operated at 200 keV, 850-nm selected-area aperture) with Protochips Fusion Select holder and Gatan OneView camera. CDW superlattice peaks are anisotropically distributed around each Bragg peak. We average the six first-order Bragg peaks and surrounding superlattice peaks to obtain the CDW diffraction pattern (Figure S11).[17]

*In situ* four-dimensional scanning transmission electron microscopy (4D-STEM) was performed on TFS Themis Z G3 (operated at 200 keV, 0.93 mrad convergence semi-angle) with DENS lightning heating and biasing holder and electron microscope pixel array detector.[70] CDW real space maps (virtual dark-field image) are formed by subtracting a background fitted from neighboring pixels and integrating intensities from all satellite peaks at each scan position. 4D-STEM provides high-resolution maps of CDW structure; however, the convergent beam localized in real space results in resolution loss in reciprocal space (diffraction peaks become discs). Smaller probe sizes in 4D-STEM require larger virtual detectors to accommodate the CDW peaks. Figure S1A shows the CDW diffraction pattern from a 1.7-nm probe wherein Bragg and superlattice peaks appear larger than transmission electron microscopy (TEM) diffraction (e.g., Figure 2B).

2D 1T-$TaS_2$ was made using an endotaxial synthesis process previously described.[17] Here, monolayers of 1T-$TaS_2$ (octahedral coordination) are embedded and isolated within bulk metallic 2H-$TaS_2$ (prismatic coordination). TEM specimens were prepared by exfoliating bulk 1T-$TaS_2$ crystals onto polydimethylsiloxane gel stamp. The sample was then transferred to TEM grids using a home-built transfer stage. 2D 1T-$TaS_2$ polytype was synthesized by heating 1T-$TaS_2$ to 720 K in high vacuum ($<10^{-7}$ Torr) for $\sim$10 min and then brought down to room temperature.[17]

## Simulating CDW melting

Hexatic CDW melting is simulated by mobile charge centers that hop according to Boltzmann statistics ($e^{-\frac{\Delta E}{kT}}$) through a Monte Carlo process. The interaction energy between charge centers is calculated using a shifted Lennard-Jones potential truncated at 100 Å.[71,72] Continuous melting processes in low dimensions occur within a wide range of interacting potentials with attractive and repulsive terms and does not require a particular Lennard-Jones potential (Figures S12 and S13). This effectively treats high-temperature CDWs as a phenomenological charge crystal. CDW amplitude collapse is modeled with the removal of charge centers at topological defect sites. The simulation is of a grand canonical ensemble (varying particle count, constant volume, and temperature). Every 150 iterations, we remove a charge center with the lowest $\Psi_6$ order parameter until we reach a minimum energy configuration. The melted CDW in Figures 3A–3F starts with a particle count of 2,016 charge centers in a fixed 2D volume with periodic boundary conditions. We ensure energy convergence at steady state. Temperature and energy were scaled to be consistent with observed 2D 1T-$TaS_2$ incommensurate CDW melting.

Diffraction of simulated CDWs in a 2D 1T-$TaS_2$ crystal is calculated using the concomitant periodic lattice displacement (PLD) of the atomic potentials. The displacement amplitude is proportional to the charge density gradient with a maximum displacement of 7 pm. Electron diffraction is kinematically simulated under flat Ewald sphere approximations using the Fourier transform of the displaced atomic lattice.

## Meta-analysis data collection

Meta-analysis data were digitally extracted from wavevector, peak width, and integrated intensity vs. temperature plots or peak profiles from digital copies of each manuscript. In some instances, the integrated intensity or peak width was not reported. References to each manuscript are included.

## RESOURCE AVAILABILITY

### Lead contact

Requests for further information and resources should be directed to and will be fulfilled by the lead contact, Robert Hovden (hovden@umich.edu).

### Materials availability

This study did not generate new materials.

### Data and code availability

- *In situ* SAED of 2D $TaS_2$ CDW is available at https://doi.org/10.5281/zenodo.15335279.
- The code used to simulate and visualize 2D CDW melting (Figures 3A–3F) is available at https://doi.org/10.5281/zenodo.15334715.

## ACKNOWLEDGMENTS

R. Hovden, A.S., and J.M.S. acknowledge support from the US Department of Energy, Basic Energy Sciences, under award DE-SC0024147. N.A., K.S., and L.Z. were supported by the National Science Foundation through the Materials Research Science and Engineering Center at the University of Michigan (award no. DMR-2309029). R. He acknowledges support by DOE Office of Science (grant no. DE-SC0020334, subaward S6535A). G.Y. was supported by the NSF (grant no. DMR-2104036). C.N. was supported by the National Science Foundation (grant no. DMR-2300640). Experiments were conducted using the Michigan Center for Materials Characterization (MC2) and MIT.nano Characterization Facilities. This research was supported in part through computational resources and services provided by Advanced Research Computing at the University of Michigan, Ann Arbor. No LLMs or AI were used to write this text.

## AUTHOR CONTRIBUTIONS

S.H.S., J.M.S., and R. Hovden analyzed electron diffraction data. J.M.S., A.S., S.H.S., I.E.B., R. Hovden, and K.S. provided theoretical interpretation. J.M.S., S.H.S., N.A., and R. Hovden ran computational models of CDW melting and diffraction. A.S. and K.S. developed analytic Landau description of CDW melting. G.Y., C.N., A.S., L.Z., and R. He performed high-temperature Raman experiments and analysis. J.M.S., Y.Z., N.A., R. Hovden, and I.E.B. conducted 4D-STEM experiments and analysis. J.M.S., A.S., K.S., and R. Hovden prepared the manuscript. All authors reviewed and edited the manuscript.

## DECLARATION OF INTERESTS

The authors declare no competing interests.

## SUPPLEMENTAL INFORMATION

Supplemental information can be found online at https://doi.org/10.1016/j.matt.2026.102665.